\documentclass[apj]{emulateapj}

\usepackage{graphicx}
\usepackage{natbib}
\usepackage{amssymb}

\begin{document}

\title{The spin parameter of uniformly rotating compact stars}

\author{Ka-Wai Lo\footnote{Current address: Department of Physics,
University of Illinois at Urbana-Champaign, Urbana, IL 61801, USA.} 
and Lap-Ming Lin\footnote{Email address for all correspondence: 
lmlin@phy.cuhk.edu.hk} }
\affil{Department of Physics and Institute of Theoretical Physics, 
The Chinese University of Hong Kong, Hong Kong, China }

\date{\today}

\begin{abstract}
We study the dimensionless spin parameter $j \ (= c J/ (G M^2) )$ of 
uniformly rotating neutron stars and quark stars in general relativity. 
We show numerically that the maximum value of the spin parameter of a neutron 
star rotating at the Keplerian frequency is $j_{\rm max} \sim 0.7$ for a wide 
class of realistic equations of state. This upper bound is insensitive to the 
mass of the neutron star if the mass of the star is larger than about 
$1\ M_\odot$. 
On the other hand, the spin parameter of a quark star modeled by 
the MIT bag model can be larger than unity and does not have a universal 
upper bound. Its value also depends strongly on the bag constant and the mass 
of the star. 
Astrophysical implications of our finding will be discussed. 
\end{abstract}


\maketitle


\section{Introduction}
\label{sec:intro}

The general stationary vacuum solution (the Kerr spacetime)
of the Einstein equations is specified uniquely by the gravitational mass 
$M$ and the angular momentum $J$ (see, e.g., \citet{wald1984}). 
If $J \le GM^2/c$, we have a rotating black hole. However, if $J > GM^2/c$, 
the Kerr spacetime would have a naked singularity without a horizon. 
One could then consider closed timelike curves and causality would be 
violated \citep{chandra1983}. While its validity has not yet been 
proven, the cosmic-censorship conjecture \citep{penrose1969} asserts that 
naked singularities cannot be formed via the gravitational collapse of a 
body. 
For this reason, it is believed that astrophysical black holes should 
satisfy the Kerr bound $j \le 1$, where $j = c J / G M^2$ is the 
dimensionless spin parameter. 
 
While the value of the spin parameter $j$ plays a fundamental role in 
black-hole physics, it appears that this is not the case for other stellar 
objects. 
In particular, there is no theoretical constraint on the value of $j$ for 
stars. 
It is known that the spin parameter of main-sequence stars depends 
sensitively on the stellar mass and can be much larger than unity
\citep{kraft1968,kraft1970,dicke1970,gray1982}. 
On the other hand, the spin parameter of compact stars has not been studied 
in detail (see below). As we shall discuss in more detail in  
Section~\ref{sec:astro}, the spin parameter of compact stars is interesting 
in its own right for two reasons. (1) It plays a role in our understanding of 
the observed quasi-periodic oscillations (QPOs) in disk-accreting compact-star 
systems. (2) It determines the final fate of the collapse of a rotating 
compact star.

Ever since the seminal work of \citet{hartle1967} who considered the limit 
of slow rotation, rotating compact stars have been studied extensively in 
general relativity. 
In the past two decades, various different numerical codes have been 
developed to construct rapidly rotating stellar models in general relativity. 
We refer the reader to \citet{stergi2003} for a review.
As the rotation frequency $f$ is a directly measurable quantity for pulsars,
it is thus reasonable that the maximum value for $f$ (i.e., the Keplerian 
frequency $f_{\rm K}$) has been one of the most studied physical quantities 
for relativistic rotating stars 
(see, e.g., \citet{cook1994,haensel1995,koranda1997,benhar2005,haensel2009}). 
However, in contrast to these previous works, we shall focus extensively on 
the spin parameter $j$.

It has been known that the spin parameter for a maximum-mass neutron star 
lies in the range $j \sim 0.6-0.7$ for most realistic equations of state
(EOS; e.g., \citet{cook1994} and \citet{salgado1994}). 
In this work, we first extend the previous works and show that the upper bound 
for the spin parameter $j_{\rm max} \sim 0.7$ is essentially independent of 
the mass of the neutron star if the gravitational mass of the star is larger 
than $\sim 1\ M_\odot$.   
Next we study the spin parameter of self-bound quark stars, which was not 
considered previously in \citet{cook1994} and \citet{salgado1994}. 
We find that the behavior of the spin parameters of neutron stars and quark 
stars is very different. 
In contrast to the case of neutron stars, the spin parameter of quark stars 
does not have a universal upper bound. 
It also depends sensitively on the parameter of the quark matter 
EOS and the mass of the star. Furthermore, the spin parameter of quark stars 
can be larger than unity. This leads us to propose that the spin 
parameter could be a useful indicator to identify rapidly rotating 
quark stars. 

The plan of this paper is as follows. Section~\ref{sec:results} presents the 
main numerical results of this work. In Section~\ref{sec:astro}, we discuss 
the astrophysical implications of our results. Our conclusions are 
summarized in Section~\ref{sec:conclude}.

\section{Numerical results}
\label{sec:results}

\subsection{Numerical method and EOS models}

We make use of the numerical code {\tt rotstar} from 
the C++ LORENE library\footnote{\tt http://www.lorene.obspm.fr/} to calculate 
uniformly rotating compact star models in general relativity. 
The code uses a multi-domain spectral method \citep{bonazzola1998} to solve 
the Einstein equations in a stationary and axisymmetric spacetime with a matter
source \citep{bonazzola1993,gourgoulhon1999}. The code has been tested 
extensively and compared with a few different numerical codes 
\citep{nozawa1998}. 

As theoretical calculations for dense matter at supranuclear 
densities are poorly constrained, the EOS in the 
high-density core of compact stars is not well understood 
(see, e.g., \citet{weber2007,haensel2007} for reviews). 
In this work, we employ eight realistic nuclear matter 
EOS to model rotating neutron stars: model A \citep{pandha1971}, 
model APR \citep{akmal1998}, model AU 
(the AV14+UVII model in \citet{wiringa1988} is joined to 
\citet{negele1973}), model BBB2 \citep{baldo1997}, 
model FPS \citep{pandha1989,lorenz1993}, 
model SLY4 \citep{douchin2000}, model UU (the UV14+UVII model in 
\citet{wiringa1988} is joined to \citet{negele1973})
and model WS (the UV14+TNI model in \citet{wiringa1988} is joined to 
\citet{lorenz1993}).
For the quark star models, we use the simplest MIT bag model with 
non-interacting massless quarks \citep{chodos1974}. 
Two different values, $60\ {\rm MeV\ fm}^{-3}$ and $90\ {\rm MeV\ fm}^{-3}$, 
are chosen for the bag constant $B$. These values correspond approximately 
to the range of $B$ within which the hypothesis of strange matter is valid 
\citep{haensel2007}. 

\begin{figure}
\begin{center}
\includegraphics*[width=8cm]{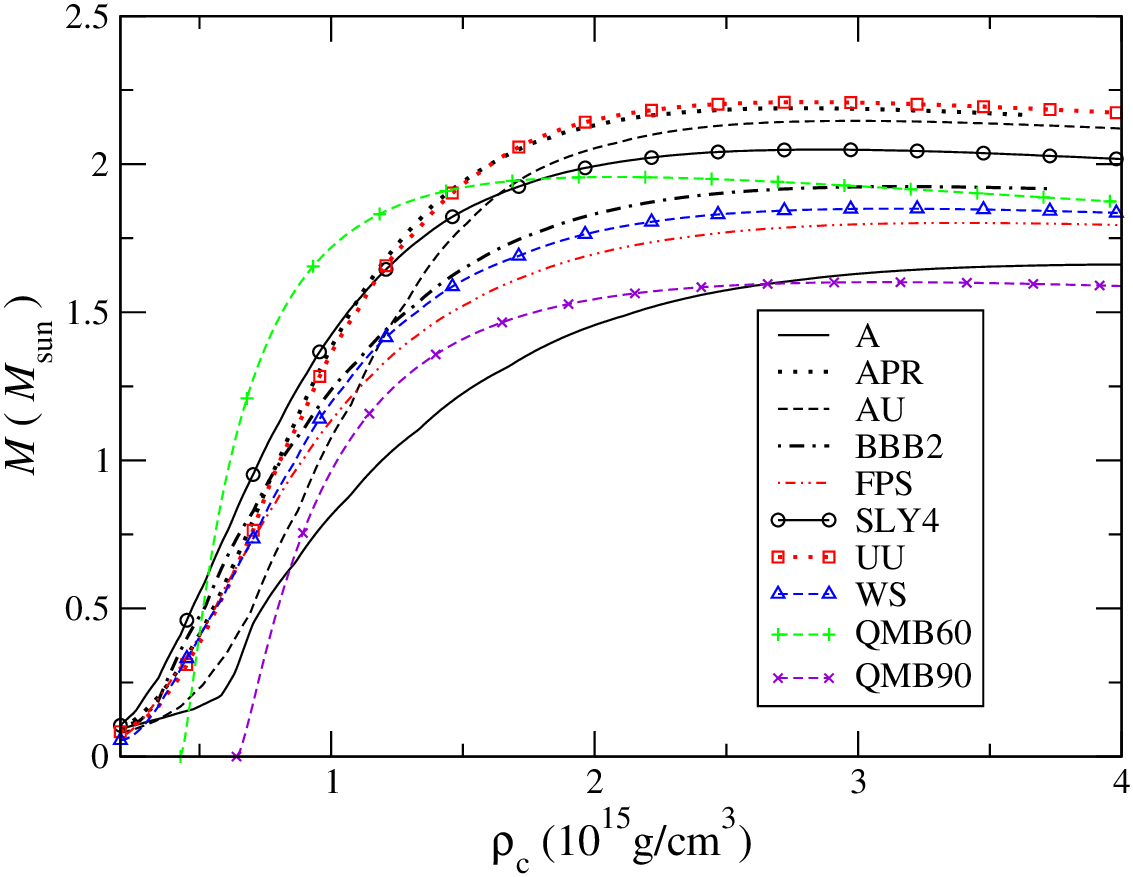}
\caption{ Gravitational mass as a function of the central 
energy density for non-rotating compact stars constructed with 
the chosen EOS models in this work.}
\label{fig:tov_curve}
\end{center}
\end{figure}

\begin{figure}
\begin{center}
\includegraphics*[width=8cm]{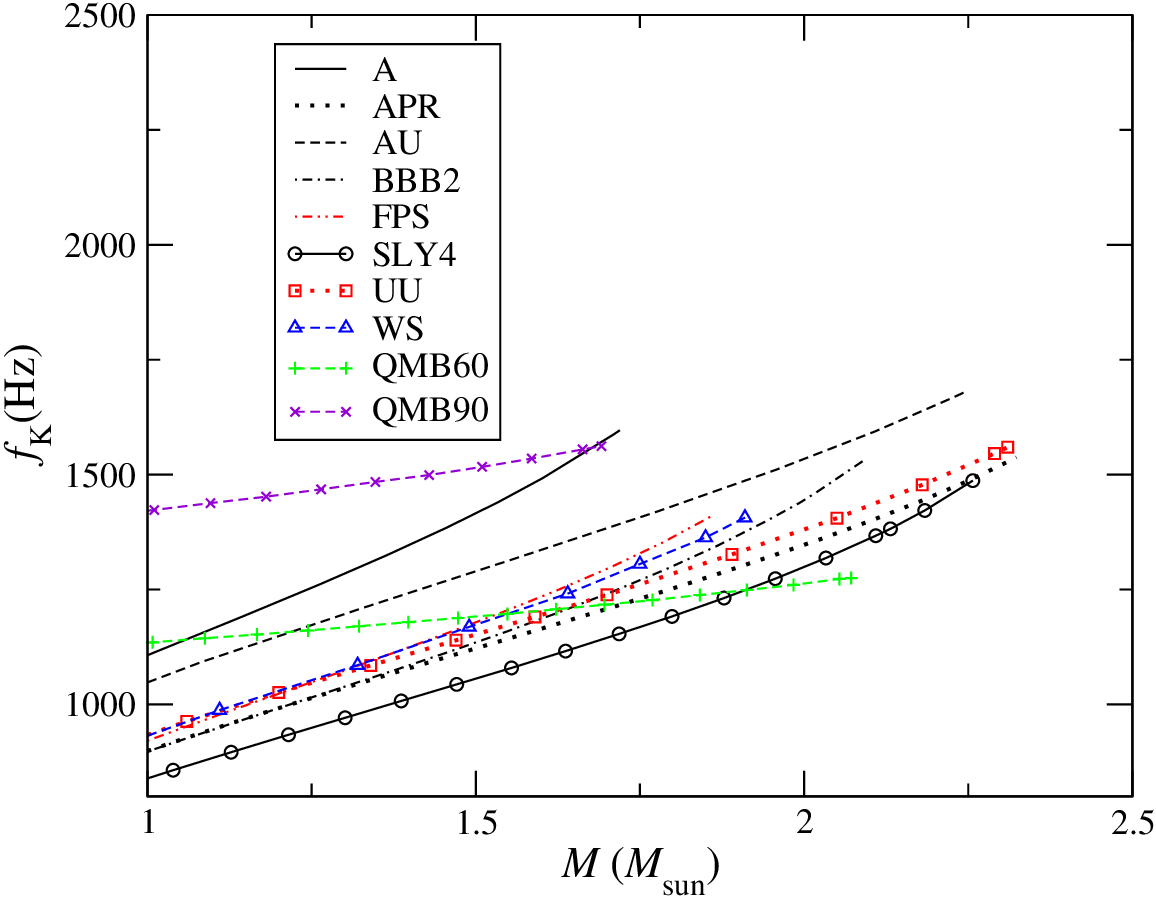}
\caption{ Keplerian frequency as a function of the gravitational 
mass for rotating compact stars constructed with the chosen EOS models in this 
work. }
\label{fig:fk_mass}
\end{center}
\end{figure}

To illustrate the diversity of the EOS models used in this work, we plot 
the gravitational mass $M$ against central energy density $\rho_{\rm c}$
for non-rotating stars constructed with the chosen EOS models in 
Figure~\ref{fig:tov_curve}. 
The quark star models QMB60 and QMB90 in the figure correspond to the 
cases $B=60\ {\rm MeV\ fm}^{-3}$ and $90\ {\rm MeV\ fm}^{-3}$, respectively. 
The maximum mass of compact stars depends quite sensitively on the EOS models
and it ranges from about $1.5\ M_{\odot}$ to $2.2\ M_{\odot}$. 
In Figure~\ref{fig:fk_mass}, we plot the Keplerian frequency
$f_{\rm K}$ against the gravitational mass $M$ of rotating compact stars 
based on our chosen EOS models. 
Similar to the maximum mass for non-rotating compact stars, 
Figure~\ref{fig:fk_mass} shows clearly that $f_{\rm K}$ depends strongly 
on the EOS model. It is also sensitive to the mass of the star. 
This is the well-known reason why searching for rapidly rotating compact stars 
can provide us constraints on the EOS of dense matter. 

\subsection{Neutron stars}

\begin{figure}
\begin{center}
\includegraphics*[width=8cm]{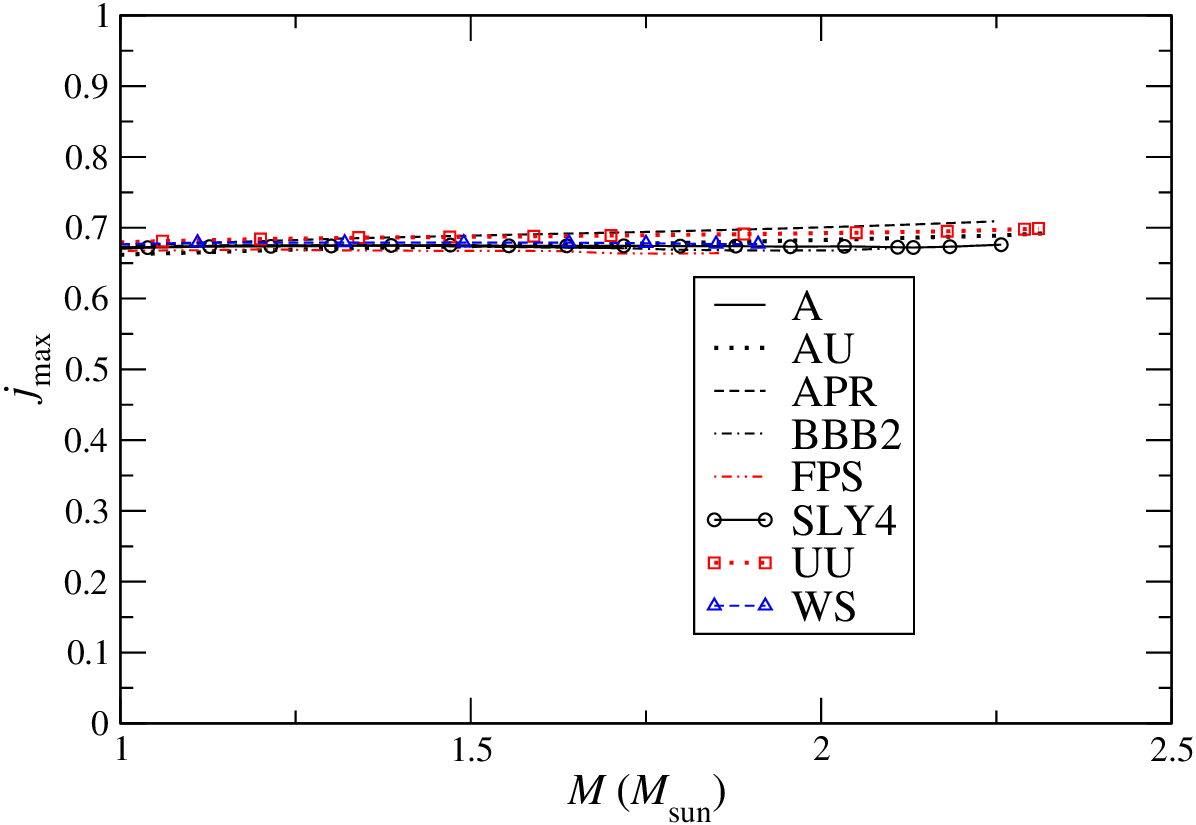}
\caption{ Maximum spin parameter as a function of the 
gravitational mass for rotating neutron stars.}
\label{fig:jmax_m2_ns}
\end{center}
\end{figure}

Now we turn to the main focus of this work: the dimensionless spin parameter
$j$. Having seen that $f_{\rm K}$ depends strongly on the EOS and the
mass of the star, it may be quite surprising to learn that the maximum value 
of the spin parameter $j_{\rm max}$ (as set by the Kepler limit) is 
quite universal for rotating neutron stars. 
In Figure~\ref{fig:jmax_m2_ns}, we plot $j_{\rm max}$ against the gravitational 
mass $M$ for the selected nuclear matter EOS models. 
In the figure, each line corresponds to one particular EOS and each point on a 
line corresponds to a star model with a fixed $M$ rotating at its 
Keplerian frequency. 
Note that each sequence in the figure is terminated at the stellar model 
with the same total particle number as the stable maximum-mass non-rotating 
configuration. 
In the figure, we see that $j_{\rm max}$ lies in a narrow range 
$\sim 0.65-0.7$ for the eight different nuclear matter EOS models. 
In particular, the values do not depend sensitively on the mass of 
the star.
This extends previous works \citep{cook1994,salgado1994}
which focus on maximum-mass neutron star models.

While the spin parameter of an astrophysical black hole is constrained by 
$j \le 1$, we find that the spin parameter of a neutron star is bounded
by $j_{\rm max} \sim 0.7$. The upper bound $j_{\rm max}$ is quite 
universal for different EOS models and gravitational mass larger than 
$\sim 1\  M_{\odot}$.  
For lower mass neutron stars, $M < 1\ M_\odot$, we find that $j$ 
decreases with decreasing $M$. However, we shall only 
focus on mass $M > 1 \ M_\odot$ in this work as the observed masses of 
neutron stars are typically larger than $1\ M_\odot$. 
We refer the reader to \citet{steiner2010} for a recent 
review on the observed masses of neutron stars.

\begin{figure}
\begin{center}
\includegraphics*[width=8cm]{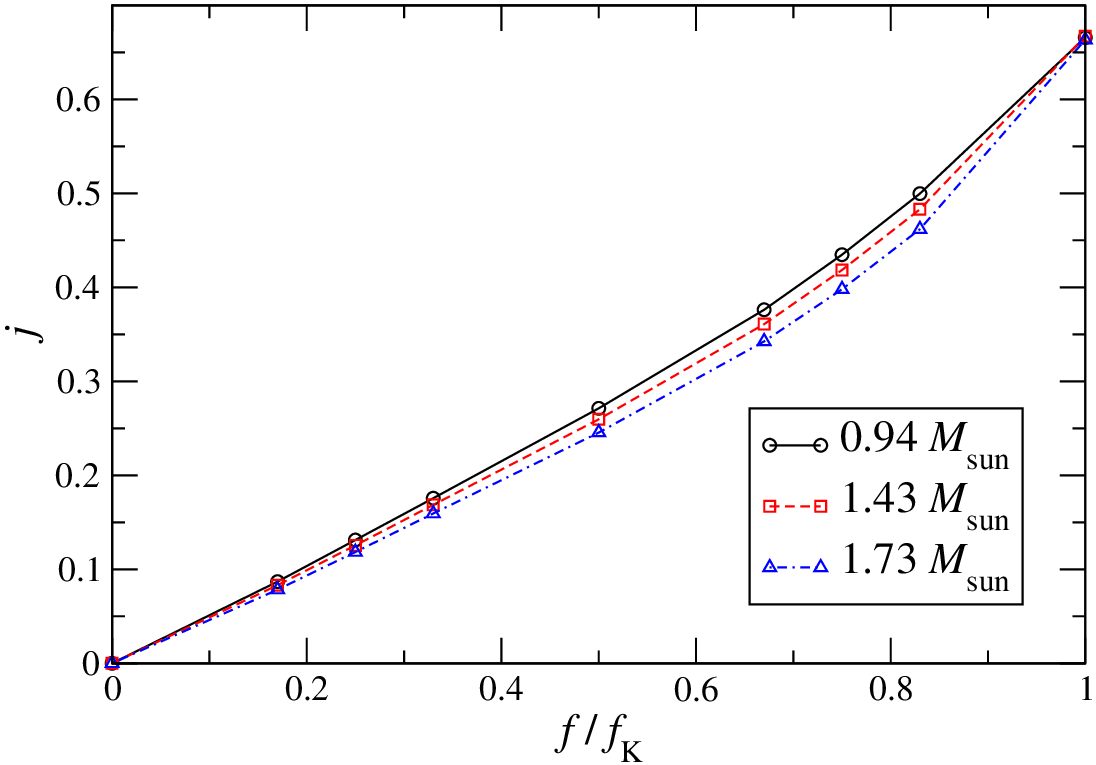}
\caption{ Spin parameter is plotted against the scaled
rotation frequency for neutron stars constructed with the FPS EOS. Each line
represents a sequence of fixed total particle number. Each sequence is 
labeled by the gravitational mass of its non-rotating configuration.  }
\label{fig:FPS_j_f}
\end{center}
\end{figure}

\begin{figure}
\begin{center}
\includegraphics*[width=8cm]{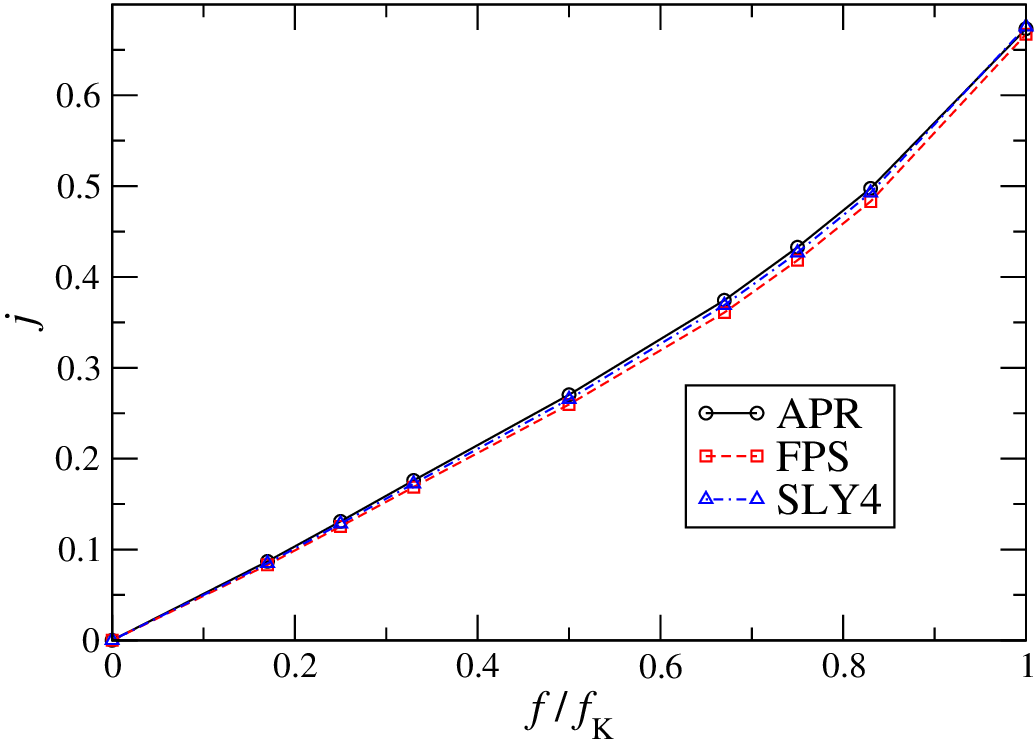}
\caption{ Spin parameter is plotted against the scaled
rotation frequency for neutron stars constructed with three different EOS 
models. The baryonic mass is fixed at $M_{\rm B} = 1.6\ M_\odot$.  }
\label{fig:NSeos_Mb1.6_j_f}
\end{center}
\end{figure}

So far we have studied the maximum spin parameter $j_{\rm max}$ of neutron
stars rotating at their Keplerian frequencies $f_{\rm K}$. 
However, realistic neutron stars in general rotate slower with frequencies 
$f < f_{\rm K}$. Is the spin parameter $j$ still insensitive to the EOS 
and mass of the star? 
In Figure~\ref{fig:FPS_j_f}, we plot the spin parameter $j$ against the scaled 
rotation frequency $f / f_{\rm K}$ for the FPS EOS. In the figure, each line 
represents a sequence of fixed total particle number (the so-called 
evolutionary sequence). 
Each sequence is labeled by the gravitational mass of its non-rotating 
configuration $M_0$. 
The solid line corresponds to $M_0 = 0.94\ M_\odot$, the dashed 
line corresponds to $M_0 = 1.43\ M_\odot$, and the dashed-dotted line
corresponds to $M_0 = 1.73\ M_\odot$. 
Note, however, that the gravitational mass increases with rotation frequency. 
The figure shows that the spin parameter changes only by at most 10\% 
(depending on $f/f_{\rm K}$) when $M_0$ changes from 0.94 to 
1.73 $M_\odot$. 
It is also interesting to note that the differences between 
the curves decrease as the scaled frequency $f / f_{\rm K}$ tends to 1. 
This is quite different from the case of quark stars which will be studied 
shortly. 
In Figure~\ref{fig:NSeos_Mb1.6_j_f}, we plot $j$ against $f/ f_{\rm K}$ 
for three different EOS models. The stellar models on the three 
sequences have the same total particle numbers such that their baryonic masses 
are fixed at $M_{\rm B} = 1.6\ M_\odot$. 
It is seen that, for a fixed scaled frequency, the spin parameter of neutron
stars is essentially independent of the EOS models.

\subsection{Quark stars}

\begin{figure}
\begin{center}
\includegraphics*[width=8cm]{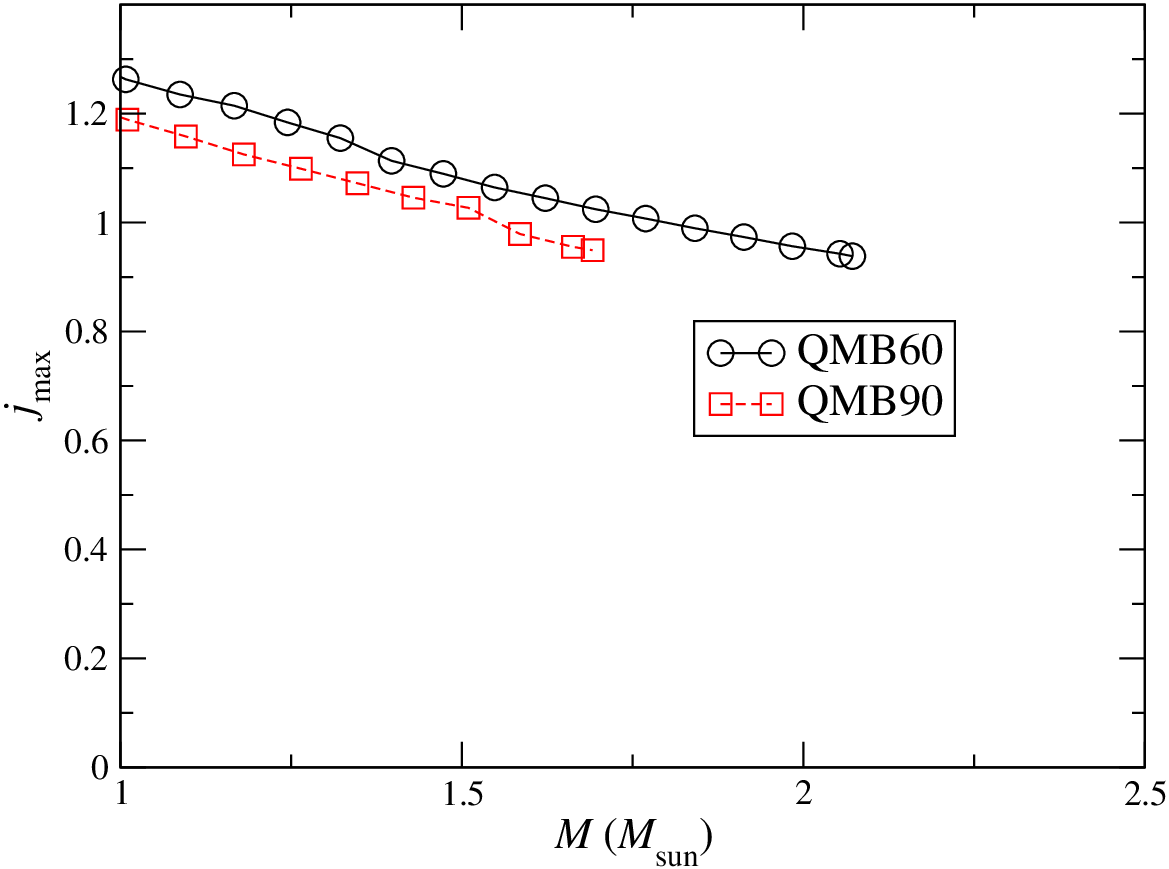}
\caption{ Maximum spin parameter as a function of the 
gravitational mass for rotating quark stars.}
\label{fig:jmax_m2_qs}
\end{center}
\end{figure}

\begin{figure}
\begin{center}
\includegraphics*[width=8cm]{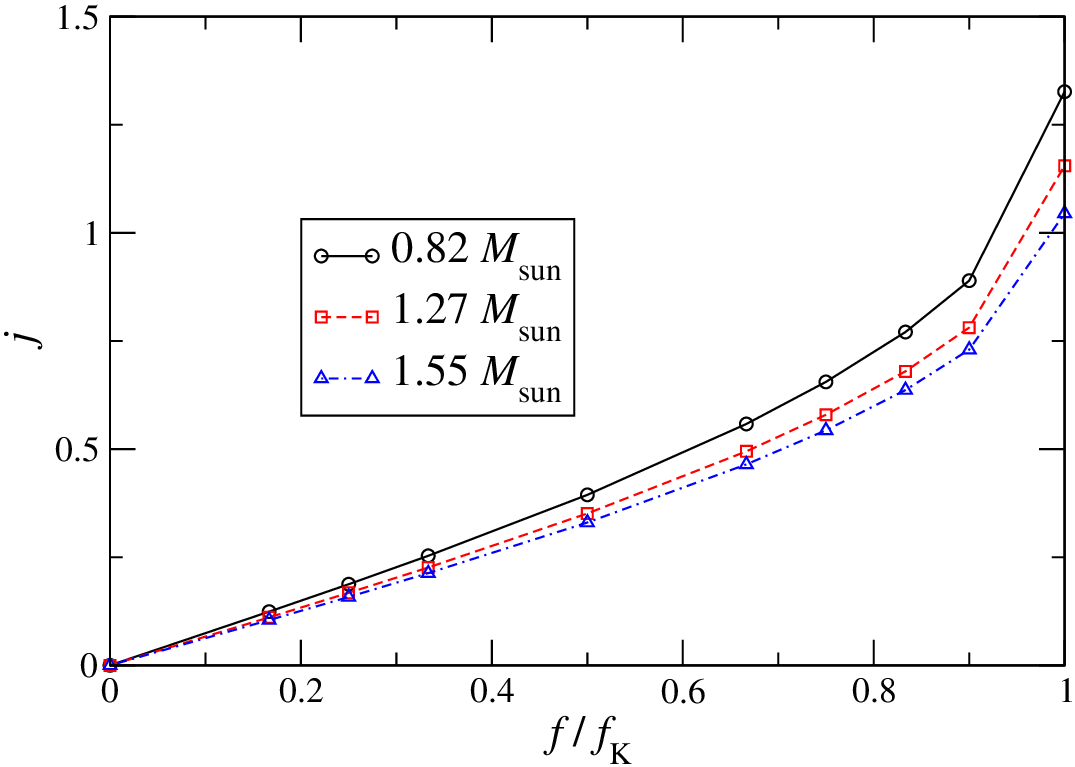}
\caption{ Spin parameter is plotted against the scaled 
rotation frequency for quark stars constructed with the QMB60 EOS. 
Each line represents a sequence of fixed total particle number. 
Each sequence is labeled by the gravitational mass of its non-rotating 
configuration. }
\label{fig:B60_j_f}
\end{center}
\end{figure}

Let us now consider self-bound quark stars using the MIT bag model. 
In Figure~\ref{fig:jmax_m2_qs}, we plot $j_{\rm max}$ against $M$ 
for the two quark matter models QMB60 and QMB90. 
In contrast to the case of neutron stars (see Figure~\ref{fig:jmax_m2_ns}), 
we see that $j_{\rm max}$ depends sensitively on the mass of the star.
For the QMB60 model, $j_{\rm max}$ is decreased by about 24\% as $M$
increases from $1\ M_\odot$ to $2\ M_\odot$. 
Comparing to the case of neutron stars, it is also seen that $j_{\rm max}$ 
has a more significant dependence on the EOS parameter, namely the bag 
constant in the MIT bag model.  
Analogous to Figure~\ref{fig:FPS_j_f} for neutron stars, 
we plot $j$ against $f / f_{\rm K}$ for the QMB60 model in 
Figure~\ref{fig:B60_j_f}. Each line represents a sequence of fixed total 
particle number. As in Figure~\ref{fig:FPS_j_f}, each sequence is labeled 
by the gravitational mass of its non-rotating configuration $M_0$. 
We show three different sequences in the figure: 
$M_0 = 0.82\ M_\odot$ (solid), 1.27 $M_\odot$ (dashed), 
and 1.55 $M_\odot$ (dash-dotted). 
We see that the differences between the curves increase significantly as 
the scaled frequency increases. 
At $f / f_{\rm K} = 1$, the spin parameter is increased by about 27\% 
when $M_0$ changes from 1.55 to 0.82 $M_\odot$.

We see that the spin parameter of quark stars can be 
significantly larger than the upper bound ($j_{\rm max} \sim 0.7$) for 
neutron stars. 
It can also be larger than the Kerr bound $j = 1$ for black holes. 
This suggests that the spin parameter could be a useful indicator to 
identify rapidly rotating quark stars. Discovering even one single compact 
star with spin parameter $j \gtrsim 0.7$ will provide a strong evidence for 
the existence of quark stars. 
Finally, it should be pointed out that the fact that the spin parameter of 
quark stars can be larger than 0.7 is also evident from the results of 
\citet{stergi1999}. 
Using the values of the gravitational mass $M$ and angular momentum $J$ for 
Keplerian quark stars presented in Table 1 of \citet{stergi1999}, it is easy 
to check that the quark stars considered by the authors all have 
$j_{\rm max} > 0.7$. In particular, their results also suggest that $j_{\max}$ 
decreases with increasing $M$ as we have seen in Figure~\ref{fig:jmax_m2_qs}.

\section{Astrophysical implications}
\label{sec:astro}

We have studied the spin parameter of uniformly rotating compact stars in
general relativity. 
Our numerical results show that the behavior of the spin parameter of quark 
stars is quite different from that of neutron stars. 
In particular, the spin parameter of neutron stars is 
bounded above by $ j_{\rm max} \approx 0.7$, while quark stars can have a 
value larger than unity. 
In this section, we shall discuss (in our view) the astrophysical 
implications of our results.

First, how could the spin parameter of a compact star be measured? 
Unfortunately, so far there is no general technique to infer the spin 
parameter $j$ of compact stars directly.
As far as we are aware, the spin parameter of a compact star could be 
potentially measured in disk-accreting compact-star systems. 
In particular, the neutron stars (or quark stars) in low-mass X-ray binaries 
(LMXBs) provide the most natural cosmic laboratories for studying the 
spin parameter. 
In order to understand how the spin parameter might be inferred in 
disk-accreting systems, it should be noted that the spin parameter of the 
central compact star directly affects the particle motion around the star. 
For example, to first order in $j$, the orbital frequency 
of a point particle in a prograde orbit around a compact star is given by 
(see, e.g., \citet{van_der_klis2006})
\begin{equation}
2 \pi \nu_{ \phi } = \left[ 1 - j \left( {G M }\over {r c^2} \right)^{3/2} 
\right] \left( {G M} \over {r^3} \right)^{1/2} \ ,
\label{eq:orbital_freq}
\end{equation}
where $r$ is the orbital radius. For infinitesimally tilted and eccentric 
orbits, the disk particles will have radial ($\nu_r$) and vertical 
($\nu_\theta$) epicyclic frequencies which also depend on $j$ 
(see \citet{van_der_klis2006} for the expressions). Furthermore, the 
combination $\nu_\theta - \nu_r$ also gives rise to the periastron frequency 
of the orbit. These frequencies in general depend on $M$ and $j$, and hence
their observations (possibly needed to be combined with the measurement
of other stellar parameters such as the mass) would in principle 
provide useful information on the spin parameter. 
In fact, there are strong evidences that these frequencies have already been 
observed in LMXBs.

It should be noted that existing algebraic relations, such as 
Equation~(\ref{eq:orbital_freq}), which relate various orbital frequencies to 
stellar parameters are in general only valid for small spin rate 
$j \sim 0.1$. The difficulty of obtaining the corresponding algebraic 
relations for rapidly rotating stars ($j \sim 1$) lies in the fact that there 
is no exact analytic representation of the vacuum spacetime outside a 
rapidly rotating compact star. Having such an analytic representation of the 
spacetime metric will allow one to obtain the desired algebraic relations for 
a rapidly rotating compact star.  
A starting point along this direction would be to take the 
closed-form asymptotically flat solution of the Einstein-Maxwell system 
obtained by \citet{manko2000} and study the geodesics in this spacetime. 
If there is no charge and magnetic moment, this analytic solution depends only 
on the mass, angular momentum, and quadrupole moment of the spacetime. 
Furthermore, 
this solution only involves rational functions, which helps to simplify the 
analytical study of geodesic motions.  
\citet{berti2004} have demonstrated that this analytic solution can 
describe the exterior spacetime of a rapidly rotating neutron star 
(e.g., $j > 0.5$) very well. Nevertheless, further investigation 
is needed to check whether this analytic solution is also valid for rapidly 
rotating quark stars.

One of the well-observed features of LMXBs is the high-frequency ($\sim$ kHz) 
QPOs. To date, QPOs have been observed in more 
than 20 LMXBs. These QPOs often come in pairs with frequencies
$\nu_u$ and $\nu_l$.
In all systems in which the spin frequencies of the compact stars 
$\nu_{\rm star}$ have been measured, the frequency separation 
$\Delta \nu = \nu_u - \nu_l$ is approximately equal to 
$\nu_{\rm star}$ or $\nu_{\rm star}/2$. We refer the reader to 
\citet{van_der_klis2006} and \citet{lamb2008} for recent reviews. 
While the physical mechanism responsible for producing the high-frequency 
QPOs is not known yet, most physical models involve orbital motion and 
disk oscillations.  Hence, the frequencies $\nu_r$, $\nu_\theta$,
$\nu_\phi$ and various combinations of them are often invoked (either 
directly or indirectly) to explain high-frequency QPOs. This is the reason 
why one might hope to obtain useful information on the spin parameter of the 
central compact star in an LMXB by observing its QPOs. 
For example, if the higher frequency of the QPO pair ($\nu_u$) is identified
with the orbital frequency ($\nu_\phi$), one can then obtain a relation 
between the spin parameter and an upper bound on the mass of the compact 
star \citep{miller1998}. 
On the other hand, based on the so-called relativistic precession model of 
QPOs \citep{stella1998,stella1999}, 
\citet{torok2010} have recently derived a constraint relating the 
mass and spin parameter of the compact star in Cir X-1. 
These works thus demonstrate the possibility of measuring the value of $j$ 
with an independent determination of the mass and vice versa.

The above brief review of high-frequency QPOs serves to point out the 
astrophysical relevance of the spin parameter and in what situations could 
it be potentially measured. 
Now we are ready to discuss how one could make use of the finding in this 
work to obtain useful information regarding the central compact star in LMXBs.
Suppose a single well established model for the high-frequency 
QPOs can be agreed upon in the future, then it is possible that rapidly 
rotating quark stars could be identified from the inferred spin parameters. 
As shown in our numerical results, if the inferred spin parameter of the 
central star is larger than $\sim 0.7$, then the star could be a quark 
star. Of course, the star can either be a neutron star or quark star if 
the spin parameter is less than 0.7.

On the other hand, for the present situation where many models are available, 
our finding could still be used to put constraints on the physical models for 
QPOs. 
For example, let us consider the resonantly excited disk-oscillation model for 
QPOs proposed by \citet{kato2008}. 
In the model, Kato suggests that the high-frequency QPOs are inertial-acoustic 
oscillations on a deformed disk that are resonantly excited by nonlinear 
couplings between the oscillation modes and the disk deformation. 
Kato applies the model to Cir X-1 and finds that it describes the observed 
QPOs 
quite well if the mass of the central star is about $1.5-2.0\ M_\odot$ and 
the spin parameter is $j \sim 0.8$. 
However, our work suggests that uniformly rotating neutron stars cannot 
have $j \gtrsim 0.7$. 
If Kato's model is correct, then our finding implies that the 
central star in Cir X-1 could be a quark star. On the other hand, if 
other measurements in the future (such as the mass-radius relation or cooling 
property of the central star) suggest that the compact star in the system 
is indeed a traditional neutron star, then our finding would rule out the
resonantly excited disk-oscillation model.

After discussing how one might use the spin parameter to distinguish between 
neutron stars and quark stars in LMXBs, let us now turn to a different issue 
concerning the collapse of a rotating star to black hole. 
In the past decade, general relativistic simulations of rotational 
collapse of neutron stars modeled by the polytropic EOS have been 
performed \citep{shibata2000a,shibata2003a,shibata2003b,duez2004,baiotti2005a}. 
These simulations show that no stable massive disks can
be formed around the resulting black holes. The implication is that the 
collapse of a uniformly rotating neutron star (with $j < 1$)  
to black hole cannot lead to the black-hole accretion model of gamma-ray 
burst, which requires a rotating black hole surrounded by an accretion 
disk (see \citet{piran2005} for a review).  
It is expected that the initial spin parameter of the
collapsing star must be $j \gtrsim 1$ in order to form a massive 
disk around the final black hole \citep{shibata2003a,shibata2003b,duez2004}. 
As rapidly rotating quark stars can have $j > 1$, it is thus possible that
the collapse of a rapidly rotating quark star could be a progenitor 
for the black-hole accretion model of gamma-ray burst. 
Furthermore, it is noted that the collapse of massive stars to black holes 
(the so-called collapsars) could form massive disks around the black holes 
and hence produce long gamma-ray bursts. 
Since quark stars have smaller mass ($\sim M_\odot$) comparing to 
massive stars, the collapse of quark stars might only produce small disks.
As the duration of the accretion, and hence the timescale of the bursts, 
depends on the mass of the disk, the collapse of quark stars might thus
be a mechanism for short gamma-ray bursts.

On the other hand, what if the collapse process does not lead to any 
(or little) mass ejection as in the collapse of neutron stars? 
In fact, \citet{bauswein2009} have recently simulated quark star mergers 
based on the MIT bag model and the conformally flat approximation to 
general relativity. 
Their results show that the mass ejection depends sensitively on the bag 
constant $B$. In particular, they find that there are some binary mergers 
without mass ejection. Could this conclusion, namely the absence of mass 
ejection for some values of $B$, also be true for the rotational collapse 
of quark stars? 
For the collapse of rotating neutron stars to black holes, apart from the 
small amount of total mass energy $M$ and angular momentum $J$ carried away by 
gravitational radiation \citep{baiotti2005b}, the final black holes have 
essentially the same $M$ and $J$, and hence the same spin parameter $j$, 
as the initial star. 
For a rapidly rotating quark star with initial spin parameter $j > 1$, 
if there is no mass ejection, how could the spin parameter be reduced 
efficiently in order to form a regular black hole that satisfies the 
Kerr bound $j \le 1$ at the end of the collapse? 
If a Kerr black hole could not be formed in the process, then what would 
be the final fate of the collapse? 
These questions deserve further investigation using fully general 
relativistic modeling. The hope is that studying the collapse of quark 
stars might lead to the discovery of some new phenomena which 
are not seen in the collapse of neutron stars. This might then help to 
shed more light on the highly nonlinear dynamics of gravitational collapse
in general relativity.

The astrophysical implications discussed above depend crucially on the 
existence of rapidly rotating quark stars with $j > 0.7$. However, it is 
known that rapidly rotating compact stars may be subject to different 
kinds of secular or dynamical non-axisymmetric mode instabilities 
\citep{andersson2003}. 
One might thus worry that rotating quark stars (if they exist) may be limited 
to small spin rates due to various instabilities, and hence rendering a 
possible distinction between neutron stars and quark stars unlikely to 
happen.   
However, \citet{gondek2003} have shown that, taking into account realistic 
values of shear viscosity, viscosity-driven bar mode instability cannot 
develop in quark stars modeled by the MIT bag model in any astrophysically 
relevant temperature windows. 
Even taking the unrealistic assumption of infinite 
shear viscosity, the instability can develop only if the ratio of the 
rotational kinetic energy to the absolute value of the gravitational potential 
energy $T/ W $ is larger than $0.1375$. The exact value depends on the stellar
mass. For comparison, a quark star 
modeled by the MIT bag model, with $B=60\ {\rm MeV\ fm}^{-3}$, 
$M=1.146\ M_\odot$, and $j=0.8$, has the value $T/ W = 0.126$. 
We have followed the same numerical procedure of \citet{gondek2003} 
to check that this quark star is indeed stable against the viscosity-driven 
instability. We refer the reader to \citet{gondek2002} and \citet{gondek2003} 
for more details on the numerical procedure. 
On the other hand, \citet{gondek2003} have also discussed that the 
gravitational-radiation driven $r$-mode instability seems to be 
unimportant for quark stars in LMXBs if the strange quark mass is 
$m_s c^2 \sim 200$ MeV (standard value) or higher. The $r$-mode instability 
can develop and may limit quark stars to small spin rates only if the strange
quark mass takes the relatively low value $m_s c^2 \sim 100$ MeV.

Even if the viscosity-driven and gravitational-radiation-driven 
instabilities (which are both secular effects) cannot develop in a quark 
star, the star may still subject to dynamical bar-mode instability which 
occurs at a higher spin rate. 
This kind of instability has not been studied for rotating quark stars. 
However, we can still obtain some insight from the Newtonian theory of a 
rotating incompressible star, which is a good approximation to quark stars
because of their rather uniform density profile.  
The onset of dynamical instability occurs at $T/ W \approx 0.27$ for an 
incompressible star \citep{chandra1969}. 
For comparison, a quark star modeled by the MIT bag 
model, with $B=60\ {\rm MeV\ fm}^{-3}$, $M = 1.4\ M_\odot$ and $j=1.11$, has 
the value $T/ W = 0.23$.
Although general relativity may change the critical value $T/W$ 
for the onset of the instability to a somewhat smaller value than that is 
suggested by the Newtonian theory\footnote{ \citet{shibata2000b} have shown 
that the effects of general relativity only change the critical value to 
$T/W \sim 0.24-0.25$ for stars modeled by a polytropic EOS.}, 
it is still very likely that rapidly rotating quark stars with $j > 0.7$ can 
exist. Some of them may even break the Kerr bound $j=1$ for black holes.

\section{Conclusions}
\label{sec:conclude}

In this paper, we have studied the dimensionless spin parameter $j$ of 
uniformly rotating neutron stars and quark stars in general relativity. 
We find that the maximum value of the spin parameter 
(as set by the Kepler limit) of neutron stars is bounded above 
by $j_{\rm max} \sim 0.7$.
This upper bound is essentially independent of the EOS of the neutron star.
It is also insensitive to the mass of the star if the mass of the 
star is larger than about $1\ M_\odot$. 
On the other hand, the spin parameter of quark stars behaves quite 
differently. 
We find that the spin parameter of quark stars modeled by the MIT bag
model can be larger than unity. It also depends sensitively on the EOS 
parameter (i.e., the bag constant) and the mass of the star.

We have discussed (in our view) the astrophysical implications of our 
finding in detail in Section~\ref{sec:astro}. We have discussed how 
the spin parameter of compact stars could be potentially measured in 
LMXBs and its relevance in the physical models for high-frequency QPOs. 
As a first application, our finding implies that the compact star in 
Cir X-1 could be a quark star if the resonantly excited 
disk-oscillation model for QPOs is correct \citep{kato2008}, since the
model requires that the spin parameter of the central star to be $j \sim 0.8$
in order to fit the observed QPOs. 
We have also speculated on how the collapse of a rotating quark star 
might be different from the collapse of a neutron star. 
As explained in Section~\ref{sec:astro}, the collapse 
of a rapidly rotating quark star with $j \gtrsim 1$ might lead to the 
formation of a stable disk around the resulting Kerr black hole. This implies
that the collapse process might form the central engine of a gamma-ray burst. 
However, a fully general relativistic dynamical calculation (which has not 
been done in this work) is required to shed light on the issue. 

In conclusion, our work suggests that discovering even one single compact star 
with spin parameter $j \gtrsim 0.7$ will provide a strong evidence for the 
existence of quark stars, and hence verifying the hypothesis 
that strange quark matter could be absolutely stable 
\citep{witten1984}.


\section*{Acknowledgments}
This work is supported by the Hong Kong Research Grants Council 
(grant no: 401807).




\begin{thebibliography}{}

\bibitem[\protect\citeauthoryear{Akmal et al.}{1998}]{akmal1998}
Akmal,~A., Pandharipande,~V.~R., \& Ravenhall,~D.~G. 1998, Phys. Rev. C,
58, 1804

\bibitem[\protect\citeauthoryear{Andersson}{2003}]{andersson2003}
Andersson,~N. 2003, Class. Quantum Grav., 20, R105

\bibitem[\protect\citeauthoryear{Baiotti et al.}{2005a}]{baiotti2005a}
Baiotti,~L., Hawke,~I., Montero,~P.~J., L\"{o}ffler,~F., Rezzolla,~L., 
Stergioulas,~N., Font,~J.~A., \& Seidel,~E. 2005a, Phys. Rev. D, 71, 024035


\bibitem[\protect\citeauthoryear{Baiotti et al.}{2005b}]{baiotti2005b}
Baiotti,~L., Hawke,~I., Rezzolla,~L., \& Schnetter,~E. 2005b, Phys. Rev. Lett.,
94, 131101


\bibitem[\protect\citeauthoryear{Baldo et al.}{1997}]{baldo1997}
Baldo,~M., Bombaci,~I., \& Burgio,~G.~F. 1997, A\&A, 328, 274

\bibitem[\protect\citeauthoryear{Bauswein et al.}{2009}]{bauswein2009}
Bauswein,~A., Janka,~H.-T., Oechslin,~R., Pagliara,~G., Sagert,~I., 
Schaffner-Bielich,~J., Hohle,~M.~M., \& Neuh\"{a}user,~R. 2009, 
Phys. Rev. Lett., 103, 011101

\bibitem[\protect\citeauthoryear{Benhar et al.}{2005}]{benhar2005}
Benhar,~O., Ferrari,~V., Gualtieri,~L., \& Marassi,~S. 2005, Phys. Rev. D, 72,
044028

\bibitem[\protect\citeauthoryear{Berti \& Stergioulas}{2004}]{berti2004}
Berti,~E., \& Stergioulas,~N. 2004, MNRAS, 350, 1416


\bibitem[\protect\citeauthoryear{Bonazzola et al.}{1998}]{bonazzola1998}
Bonazzola,~S., Gourgoulhon,~E., \& Marck,~J.-A. 1998, Phys. Rev. D, 58, 104020


\bibitem[\protect\citeauthoryear{Bonazzola et al.}{1993}]{bonazzola1993}
Bonazzola,~S., Gourgoulhon,~E., Salgado,~M., \& Marck,~J.-A. 1993, A\&A, 278, 421

\bibitem[\protect\citeauthoryear{Chandrasekhar}{1969}]{chandra1969}
Chandrasekhar,~S. 1969, Ellipsoidal Figures of Equilibrium (New Haven, CT: Yale
Univ. Press)

\bibitem[\protect\citeauthoryear{Chandrasekhar}{1983}]{chandra1983}
Chandrasekhar,~S. 1983, The Mathematical Theory of Black Holes (New York:
Oxford Univ. Press)


\bibitem[\protect\citeauthoryear{Chodos et al.}{1974}]{chodos1974}
Chodos,~A., Jaffe,~R.~L., Johnson,~K., Thorn,~C.~B., \& Weisskopf,~V.~F. 1974,
Phys. Rev. D, 9, 3471

\bibitem[\protect\citeauthoryear{Cook et al.}{1994}]{cook1994}
Cook,~G., Shapiro,~S.~L., \& Teukolsky,~S.~A. 1994, ApJ, 424, 823


\bibitem[\protect\citeauthoryear{Dicke}{1970}]{dicke1970}
Dicke,~R.~H. 1970, in Stellar Rotation, ed. A. Slettebak 
(Dordrecht: Reidel), 289

\bibitem[\protect\citeauthoryear{Douchin \& Haensel}{2000}]{douchin2000}
Douchin,~F., \& Haensel,~P. 2000, Phys. Lett. B 485, 107

\bibitem[\protect\citeauthoryear{Duez et al.}{2004}]{duez2004}
Duez,~M.~D., Shapiro,~S.~L., \& Yo,~H.-J. 2004, Phys. Rev. D, 69, 104016


\bibitem[\protect\citeauthoryear{Gondek-Rosi\'{n}ska \& Gourgoulhon}{2002}]{gondek2002}
Gondek-Rosi\'{n}ska,~D., \& Gourgoulhon,~E. 2002, Phys. Rev. D, 66, 044021

\bibitem[\protect\citeauthoryear{Gondek-Rosi\'{n}ska et al.}{2003}]{gondek2003}
Gondek-Rosi\'{n}ska,~D., Gourgoulhon,~E., \& Haensel,~P. 2003, A\&A, 412, 777


\bibitem[\protect\citeauthoryear{Gourgoulhon et al.}{1999}]{gourgoulhon1999}
Gourgoulhon,~E., Haensel,~P., Livine,~R., Paluch,~E., Bonazzola,~S., \& Marck,~J.-A. 1999, A\&A, 349, 851

\bibitem[\protect\citeauthoryear{Gray}{1982}]{gray1982}
Gray,~D.~F. 1982, ApJ, 261, 259


\bibitem[\protect\citeauthoryear{Haensel et al.}{2007}]{haensel2007}
Haensel,~P., Potekhin,~A.~Y., \& Yakovlev,~D.~G. 2007, Neutron Stars. 1. Equation of State and Structure (New York: Springer)


\bibitem[\protect\citeauthoryear{Haensel et al.}{1995}]{haensel1995}
Haensel,~P., Salgado,~M., \& Bonazzola,~S. 1995, A\&A, 296, 746


\bibitem[\protect\citeauthoryear{Haensel et al.}{2009}]{haensel2009}
Haensel,~P., Zdunik,~J.~L., Bejger,~M., \& Lattimer,~J.~M. 2009, A\&A, 502, 605


\bibitem[\protect\citeauthoryear{Hartle}{1967}]{hartle1967}
Hartle,~J.~B. 1967, ApJ, 150, 1005


\bibitem[\protect\citeauthoryear{Kato}{2008}]{kato2008}
Kato,~S. 2008, PASJ, 60, 889

\bibitem[\protect\citeauthoryear{Koranda et al.}{1997}]{koranda1997}
Koranda,~S., Stergioulas,~N., \& Friedman,~J.~L. 1997, ApJ, 488, 799

\bibitem[\protect\citeauthoryear{Kraft}{1968}]{kraft1968}
Kraft,~R.~P. 1968, in Stellar Astronomy, ed. H.~U. Chiu et al. (New York:
Gordon \& Breach), 317

\bibitem[\protect\citeauthoryear{Kraft}{1970}]{kraft1970}
Kraft,~R.~P. 1970, in Spectroscopic Astrophysics, ed. G.~H. Herbig
(Berkeley, CA: Univ. California Press), 385 

\bibitem[\protect\citeauthoryear{Lamb \& Boutloukos}{2008}]{lamb2008}
Lamb,~F.~K., \& Boutloukos,~S. 2008, in Short-Period Binary Stars: 
Observations, Analyses, and Results, ed. E.~F. Milone et al. 
(Berlin: Springer), 87


\bibitem[\protect\citeauthoryear{Lorenz et al.}{1993}]{lorenz1993}
Lorenz,~C.~P., Ravenhall,~D.~G., \& Pethick,~C.~J. 1993, Phys. Rev. Lett., 70, 379


\bibitem[\protect\citeauthoryear{Manko et al.}{2000}]{manko2000}
Manko,~V.~S., Mielke,~E.~W., \& Sanabria-G\'{o}mez,~J.~D. 2000, 
Phys. Rev. D, 61, 081501


\bibitem[\protect\citeauthoryear{Miller et al.}{1998}]{miller1998}
Miller,~M.~C., Lamb,~F.~K., \& Psaltis,~D. 1998, ApJ, 508, 791


\bibitem[\protect\citeauthoryear{Negele \& Vautherin}{1973}]{negele1973}
Negele,~J.~W., \& Vautherin,~D. 1973, Nucl. Phys. A, 207, 298

\bibitem[\protect\citeauthoryear{Nozawa et al.}{1998}]{nozawa1998}
Nozawa,~T., Stergioulas,~N., Gourgoulhon,~E., \& Eriguchi,~Y. 1998, A\&AS, 132, 431

\bibitem[\protect\citeauthoryear{Pandharipande}{1971}]{pandha1971}
Pandharipande,~V. 1971, Nucl. Phys. A, 174, 641

\bibitem[\protect\citeauthoryear{Pandharipande \& Ravenhall}{1989}]{pandha1989}
Pandharipande,~V., \& Ravenhall,~D.~G. 1989, in Proc. NATO Advanced
Research Workshop on Nuclear Matter and Heavy Ion Collisions, Les Houches, 
ed. M. Soyeur et al. (New York: Plenum), 103

\bibitem[\protect\citeauthoryear{Penrose}{1969}]{penrose1969}
Penrose,~R. 1969, Riv. Nuovo Cimento, 1, 252
[Gen. Rel. Grav., 34, 1141 (2002)] 

\bibitem[\protect\citeauthoryear{Piran}{2005}]{piran2005}
Piran,~T. 2005, Rev. Mod. Phys., 76, 1143

\bibitem[\protect\citeauthoryear{Salgado et al.}{1994}]{salgado1994}
Salgado,~M., Bonazzola,~S., Gourgoulhon,~E., \& Haensel,~P. 1994, A\&AS, 108, 455

\bibitem[\protect\citeauthoryear{Shibata}{2003a}]{shibata2003a}
Shibata,~M. 2003a, Phys. Rev. D, 67, 024033

\bibitem[\protect\citeauthoryear{Shibata}{2003b}]{shibata2003b}
Shibata,~M. 2003b, ApJ, 595, 992


\bibitem[\protect\citeauthoryear{Shibata et al.}{2000a}]{shibata2000a}
Shibata,~M., Baumgarte,~T.~W., \& Shapiro,~S.~L. 2000a, Phys. Rev. D, 61, 044012


\bibitem[\protect\citeauthoryear{Shibata et al.}{2000b}]{shibata2000b}
Shibata,~M., Baumgarte,~T.~W., \& Shapiro,~S.~L. 2000b, ApJ, 542, 453


\bibitem[\protect\citeauthoryear{Steiner et al.}{2010}]{steiner2010}
Steiner,~A.~W., Lattimer,~J.~M., \& Brown,~E.~F. 2010, ApJ, 722, 33 


\bibitem[\protect\citeauthoryear{Stella \& Vietri}{1998}]{stella1998}
Stella,~L., \& Vietri,~M. 1998, ApJ, 492, L59

\bibitem[\protect\citeauthoryear{Stella \& Vietri}{1999}]{stella1999}
Stella,~L., \& Vietri,~M. 1999, Phys. Rev. Lett., 82, 17


\bibitem[\protect\citeauthoryear{Stergioulas}{2003}]{stergi2003}
Stergioulas,~N., 2003, Living Rev. Rel., 6, 3

\bibitem[\protect\citeauthoryear{Stergioulas et al.}{1999}]{stergi1999}
Stergioulas,~N., Klu\'{z}niak,~W., \& Bulik,~T. 1999, A\&A, 352, L116


\bibitem[\protect\citeauthoryear{T\"{o}r\"{o}k et al.}{2010}]{torok2010}
T\"{o}r\"{o}k,~G., Bakala,~P., \v{S}r\'{a}mkov\'{a},~E., Stuchl\'{i}k,~Z.,
\& Urbanec,~M. 2010, ApJ, 714, 748 


\bibitem[\protect\citeauthoryear{van~der~Klis}{2006}]{van_der_klis2006}
van~der~Klis,~M. 2006, in Compact Stellar X-ray Sources, ed. W. Lewin
\& M. van der Klis (Cambridge: Cambridge Univ. Press), 39

\bibitem[\protect\citeauthoryear{Wald}{1984}]{wald1984}
Wald,~R.~M. 1984, General Relativity (Chicago, IL: Univ. Chicago Press)

\bibitem[\protect\citeauthoryear{Weber et al.}{2007}]{weber2007}
Weber,~F., Negreiros,~R., \& Rosenfield,~P. 2007, arXiv:0705.2708v2

\bibitem[\protect\citeauthoryear{Wiringa et al.}{1988}]{wiringa1988}
Wiringa,~R.~B., Fiks,~V., \& Fabrocini,~A. 1988, Phys. Rev. C, 38, 1010


\bibitem[\protect\citeauthoryear{Witten}{1984}]{witten1984}
Witten,~E. 1984, Phys. Rev. D, 30, 272


\end{thebibliography}
\end{document}